\documentclass[10pt,letterpaper]{article} 
\usepackage[english]{babel}
\newcommand{\lb}{\langle}
\newcommand{\rb}{\rangle}

\newcommand{\beq}{\begin{equation}}
\newcommand{\eeq}{\end{equation}}
\newcommand{\lbl}{\label}
\newcommand{\beqnar}{\begin{eqnarray}}
\newcommand{\eeqnar}{\end{eqnarray}}
\newcommand{\beqnars}{\begin{eqnarray*}}
\newcommand{\eeqnars}{\end{eqnarray*}}

\newcommand{\goesto}{\rightarrow}
\newcommand{\s}{\\[1ex]}
\newcommand{\re}[1]{(\ref{#1})}
\newcommand{\q}{\quad}

\newcommand{\tr}{\mbox{Tr }}

\newcommand{\dajA}{\mbox{\boldmath $\hat{A}$}} 
\newcommand{\dajrho}{\mbox{\boldmath $\hat{\rho}$}} 
\newcommand{\dajE}{\mbox{\boldmath $\hat{E}$}}
\newcommand{\dajM}{\mbox{\boldmath $\hat{M}$}}
\newcommand{\strho}{\rho}
\newcommand{\stA}{A}
\newcommand{\stM}{M}
\newcommand{\stP}{P}
 
\newcommand{\stE}{E}

\begin{document} 
\title{ 
Counterexample to ``Sufficient conditions for uniqueness of the Weak
Value'' by J.\ Dressel and A.\ N.\ Jordan, arXiv:1106.1871v1.
} 
\author{Stephen Parrott\thanks{For contact information, 
go to http://www.math.umb.edu/$\sim$sp}} 
\date{June 24, 2011}
\maketitle 
\begin{abstract}The abstract of ``Contextual Values of Observables
in Quantum Measurements'' by J.\ Dressel, S.\ Agarwal, and A. N. Jordan 
[Phys. Rev. Lett. {\bf104} 240401 (2010)] (called DAJ
below),  states: 
\begin{quote} 
``We introduce contextual values as a generalization of the 
eigenvalues of an observable that takes into account both
the system observable and a general measurement procedure.  This 
technique leads to a natural definition of a general conditioned
average that converges uniquely to the quantum weak value in 
the minimal disturbance limit.''
\end{quote}
A counterexample to the claim of the last sentence was presented in 
\cite{parrott3}, a 32-page paper discussing various topics related 
to DAJ, and a simpler counterexample in Version 1 of the present work.  
Subsequently Dressel and Jordan placed in the arXiv the paper of the
title (called DJ below) which attempts to prove the claim of DAJ quoted
above under stronger hypotheses than given in DAJ, hypotheses which the 
counterexample does not satisfy. The present work (Version 5) presents a new 
counterexample to this claim of DJ.
\s
A brief introduction to ``contextual values'' is included.  Also included 
is a critical analysis of DJ.  
\end{abstract} 
\section{Introduction} 
A counterexample to a major claim of
\begin{quote}
J. Dressel, S Agarwal, and A.\ N.\ Jordan, ``Contextual values
of observables in quantum measurements'', Phys.\ Rev.\ Lett.\ {\bf 104}
240401 (2010)
\end{quote}
(henceforth called DAJ)
was given in \cite{parrott3}, a 32-page paper discussing DAJ in detail.
The claim in question is stated as follows in DAJ's abstract:
\begin{quote}
``We introduce contextual values as a generalization of the 
eigenvalues of an observable that takes into account both
the system observable and a general measurement procedure.  This 
technique leads to a natural definition of a general conditioned
average that converges uniquely to the quantum weak value in 
the minimal disturbance limit.''
\end{quote}
This wording (particularly, ``minimal disturbance limit'') is potentially 
misleading, as will be explained briefly below, and is discussed 
more fully in \cite{parrott3}.

Version 1 presented a simple counterexample to the claim of the above quote
based on my interpretation of the vague presentation of DAJ.  
A later paper by Dressel and Jordan, 
``Sufficient conditions for uniqueness of the Weak Value''
\cite{DJ} (henceforth called DJ) adjoined new (and very strong) hypotheses
to DAJ which the counterexample did not satisfy and claimed to prove
that the above quote was correct under these new hypotheses. 

The present work presents a new counterexample to that claim.  
It also includes the introduction to the 
main ideas of DAJ of Version 1 and a critical analysis of DJ.
\section{Notation and brief reprise of DAJ}

To establish notation, we briefly summmarize the main ideas of DAJ.
The notation generally follows DAJ except that DAJ denotes operators
by both boldface and circumflex, e.g., $\dajM$, but we omit the boldface
and ``hat'' decorations.  Also, we use $P_f$ to denote the operator of 
projection onto the subspace spanned by a vector $f$. (DAJ uses
{\boldmath $\hat{E}^{(2)}_f$}.)   

When we quote directly an equation of DAJ, we use DAJ's equation number,
which ranges from (1) to (10),  
and also DAJ's original notation.  
Other equations will bear numbers beginning
with (100). 

Suppose we are given a set $\{M_j\}$ of measurement operators, 
where $j$ is an index ranging over a finite set.
We assume that the reader is familiar with the theory of measurement operators,
as given, for example, in the book \cite{N/C} of Nielsen and Chuang.
By definition, measurement operators satisfy
\setcounter{equation}{99}
\beq
\lbl{eq100}
\sum_j M^\dag_j M_j = I \q,
\eeq
where $I$ denotes the identity operator.  With such measurement operators
is associated the {\em positive operator valued measure} (POVM)
$\{E_j\}$ with $E_j := M^\dag_j M_j$.  When the system is in a 
(generally mixed) normalized state $\rho$ 
(represented as a positive operator of 
trace 1), the probability of a measurement yielding result $j$ is 
$\tr [  M^\dag_j M_j \rho ] = \tr [ E_j \rho ]$.  Moreover, after the 
measurement, the system will be in (unnormalized) state
$M_j \rho M^\dag_j$, which when normalized is: 
\beq
\lbl{eq105} 
\mbox{normalized post-measurement state} = 
\frac{M_j \rho M^\dag_j}{\tr [M_j \rho M^\dag_j ]} 
\q.
\eeq
For notational simplicity, we normalize states only in calculations 
where the normalization factor is material.

We also assume given an operator $A$, representing what DAJ calls 
``the system observable'' in the above quote.  We ask if it is possible
to choose real numbers $\alpha_j$, which DAJ calls {\em contextual values},
such that
\beq
\lbl{eq110}
A = \sum_j \alpha_j E_j \q.
\eeq
This will not always be possible, but we consider only cases for which
it is.  When it is possible, it follows that the expectation $\tr[A\rho]$
of $A$ in the state $\rho$ equals the expectation calculated from the 
probabilities $\tr [E_j \rho ]$ obtained from the POVM $\{E_j\}$, 
with the numerical value $\alpha_j$ associated with outcome
$j$:
\beq
\lbl{eq120}
\tr [A\rho] = \sum_j \alpha_j \tr [E_j \rho]
\q.
\eeq

The book \cite{wiseman} of Wiseman and Milburn defines a measurement to 
be  ``minimally disturbing'' if the measurement operators $M_j$ are all
positive (which implies that they are Hermitian).%
\footnote{This is a technical
definition which can be misleading if one does not realize that normal
associations of the English phrase ``minimally disturbing'' are
not implied.  Further discussion can be found in \cite{wiseman} and 
\cite{parrott3}.}  
DAJ uses a slightly more general definition to define their 
``minimal disturbance limit'' of the above quote.  We shall use
the definition of Wiseman and Milburn \cite{wiseman} because it is simpler
and  sufficient for our counterexample.  A counterexample under the 
definition of Wiseman and Milburn will also be a counterexample under any more
inclusive definition, such as that of DAJ.  

A particularly simple kind of measurement is one in which there are 
only two measurement operators, $P_f$ and $I-P_f$.  Intuitively, this 
``measurement'' asks whether the (unnormalized) 
post-measurement state is $P_f$ or not.
Here we are using the notation of mixed states.  Phrased in terms
of pure states, and assuming that the pre-measurement state $\rho$ is pure, 
the measurement determines if the post-measurement state is the pure
state $f$ or a pure state orthogonal to $f$. 

Suppose that we make a measurement with the original measurement operators
$M_j$ and then make a second measurement with measurement operators
$P_f, I-P_f$.  In this situation, the second measurement is called a 
``postselection'', and when it yields state $P_f$, one says that the 
postselection has been ``successful''.  

Such a compound measurement may be 
equivalently considered as a single measurement with measurement 
operators $\{P_fM_j, (I-P_f)M_j\}$.  
``Successful'' postselection leaves the system in normalized state
\beq
\lbl{eq130} 
\frac{(P_f M_j)\rho (P_f M_j)^\dag}
{\tr [ (P_f M_j)\rho (P_f M_j)^\dag]}
\q,
\eeq
which is pure state $f$ ($P_f$ in mixed state notation).  This result will
occur with probability $p(j,f) = \tr [ (P_f M_j)^\dag P_f M_j\rho ] = 
\tr[M^\dag_j P_f M_j \rho ] .$ 

The probability $p(j|f)$ of first measurement result $j$ given that 
the postselection was successful is: 
\beq
\lbl{eq140}
p(j|f) = \frac{p(j,f)}{\sum_i p(i,f)} = 
\frac{\tr [ M^\dag_j P_f M_j \rho ]}
{\sum_i \tr [ M^\dag_i P_f M_i \rho]}
\q.
\eeq
Hence, if we assign numerical value $\alpha_j$ to result $j$ as above,
the conditional expectation of the measurement {\em given} successful
postselection is:
\beq
\lbl{eq150}
_f\lb A \rb := 
\frac{ \sum_j \alpha_j \tr [ M^\dag_j P_f M_j \rho ]}
{\sum_i \tr [ M^\dag_i P_f M_i \rho]}
\q.
\eeq
This is DAJ's ``general conditioned average''.  
Written in DAJ's original notation, this reads 
$$ 
\lbl{eq6}
_f \lb {\cal A} \rb = \sum_j \alpha^{(1)}_j P_{j|f}
= 
\frac{\sum_j \alpha^{(1)}_j \tr [ 
\hat{\bf E}^{(1,2)}_{jf} \dajrho]
}
{\sum_j \tr [ 
\hat{\bf E}^{(1,2)}_{jf} \dajrho]
}.
\eqno (6) 
$$ 

DAJ's theory of contextual values was motivated by a 
theory of ``weak measurements'' initiated by Aharonov, Albert, and Vaidman
\cite{AAV} in 1988.  
Intuitively, a ``weak'' measurement is one which negligibly
disturbs the state of the system.  This can be formalized by introducing
a ``weak measurement'' parameter $g$ on which the measurement operators
$M_j = M_j(g)$ depend, and requiring that 
\beq
\lbl{eq160} 
\lim_{g \goesto 0} 
\frac{M_j (g) \rho M^\dag_j (g) }
{\tr [M_j (g) \rho  M^\dag_j (g) ]} 
= \rho 
\q \mbox{ for all $\rho$ and $j$} 
\q,
\eeq 
This says that for small $g$, the post-measurement state is almost the
same as the pre-measurement state $\rho$ (cf. equation \re{eq130}).  
We shall refer to this as 
``weak measurement'' or a ``weak limit''.   

The ``minimal disturbance limit'' mentioned in the above quote from DAJ's
abstract presumably refers to \re{eq160} combined with their generalization
of Wiseman and Milburn's ``minimally disturbing'' condition that
the measurement operators be positive, and this is the definition that
we shall use.%
\footnote{DAJ only partially and unclearly defines 
its ``minimally disturbing'' condition,
but in a message to Physical Review Letters (PRL) in response to a 
``Comment'' paper that I submitted, the authors of DAJ confirmed that Wiseman
and Milburn's definition implies theirs.  DAJ uses but does not define 
the phrase ``weak limit'', but in the same message to PRL, the authors
state that \re{eq160} corresponds to ``ideally weak measurement''.
Since ``ideally weak measurement'' must be (assuming normal usage of syntax) 
a special case of mere ``weak measurement'', our counterexample which 
assumes \re{eq160} will also be a counterexample to the statement of DAJ
quoted in the introduction.  
\s
I have made several direct inquiries to the authors of DAJ requesting 
a precise definition
of their ``minimal disturbance limit'', but all have been ignored.
} 

DAJ claims that in their ``minimal disturbance limit''
(which is implied by a weak limit with positive  
measurement operators), their
``general conditioned average'' $_f\lb A \rb$ (6), our \re{eq150}, 
is always given by:
\beq
\lbl{eq170}
_f \lb A \rb = 
\frac{1/2 \tr [ P_f \{A, \rho\}]]}
{\tr  [P_f \rho ]}
\q.  
\eeq 
Our equation \re{eq170} is equation (7) of DAJ: 
$$
\lbl{eq7} 
A_w = \frac{\tr [\dajE^{(2)}_f \{ \dajA , \dajrho \} ]} 
{2 \tr [\dajE^{(2)}_f \dajrho ]} \q, 
\eqno (7)
$$ 
Here $A_w$ is their notation for ``weak value'' of $A$.%
\footnote
{In the traditional theory of ``weak measurement'' 
initiated by \cite{AAV}, the weak limit (i.e., $\lim_{g \goesto 0}$) of
\re{eq150} (equivalently, (6)) would be called a ``weak value'' of $A$, 
though the traditional
``weak measurement'' literature calculates it via different procedures.  
When $\rho$ is a pure state, most modern authors calculate this weak value
as \re{eq170} (equivalently (7)), 
though the seminal paper \cite{AAV} arrived (via questionable
mathematics) at a complex
weak value of which \re{eq170} is the real part. 
(Only recently was it recognized that ``weak values'' are not unique
\cite{jozsa}\cite{parrott1}\cite{parrott2}.)
}

The statement of DAJ quoted in the Introduction, that their 
\begin{quote}
``$\ldots$ general conditioned
average $\ldots$ converges uniquely to the quantum weak value in 
the minimal disturbance limit'',
\end{quote}
implies that for a weak limit of positive measurement operators, their
(6) always evaluates to (7), or in our notation, our \re{eq150} always
evaluates to \re{eq170}.  We shall give an example for which 
\re{eq150} does {\em not} evaluate to \re{eq170}.  
\section{The counterexample}
\subsection{General discussion}

We are assuming the ``minimal disturbance'' condition that 
the measurement operators be positive, so in the definition \re{eq150} 
of DAJ's ``general conditioned average'', we replace $\stM^\dag_j$ with
$\stM_j$. First we examine its denominator. 

Let
\beq
\lbl{eq180}
\eta_j(g) := \tr [ \stM_j(g) \strho\stM_j(g)  ] 
\q,
\eeq
which are inverse normalization factors for the unnormalized post-measurement
states $ \stM_i(g) \strho\stM_i(g) $ (cf. \re{eq105}.
We shall assume that all $\eta_j (g)$ are bounded for small $g$,
which is expected (because we expect $M_j(g)$ to approach a multiple 
of the identity for small $g$ in order to make the measurement
``weak'')  and will be the case for our counterexample. 
We have
\begin{eqnarray} 
\lbl{eq190} 
\lefteqn{
\lim_{g \goesto 0}
\sum_j \tr [\stP_f \stM_j (g) \strho \stM_j (g)] 
=
} 
\nonumber\\ 
&&
\lim_{g \goesto 0} \sum_j \tr [ P_f 
\left( 
\frac{ 
\stM_j (g) \strho  \stM_j (g) 
}
{\eta_j(g) } 
- \strho
\right)  
]
\,
\eta_j(g)
 \nonumber 
 \\ 
&& 
\q\q\q\q\q + \lim_{g \goesto 0} 
\sum_j \tr[\stP_f\strho ] \,  \eta_j(g) \nonumber\\
&=& 
\lim_{g \goesto 0}
\sum_j \tr[\stP_f\strho ]\, \tr [\stM_j (g) \strho  \stM_j (g) ] 
\nonumber \\
&=& 
\tr [\stP_f\strho] \lim_{g \goesto 0} \tr [ \sum_j  
\stM_j (g) \stM_j (g) \strho ]   
\nonumber \\
&=& \tr [\stP_f \strho] \q,
\end{eqnarray}
because $\sum M^2_j = \sum M^\dag_j M_j = I$ and $\tr [ \strho ] = 1.$.
This is the denominator of DAJ's claimed result \re{eq170} 
(half the denominator of their (7) because both numerator and denominator
of our \re{eq170} differ from (7) by a factor of 1/2). 

Next we examine the numerator of the ``general conditioned average'' 
\re{eq150}.  We shall write it as a sum of two terms, the first term leading
to DAJ's \re{eq170}, and the second 
a term which does not obviously vanish in the limit
$g \goesto 0.$.  
The counterexample will be obtained by finding a case for which the 
limit of the 
second term actually does not vanish.

Note the trivial identity for operators $\stM, \strho$: 
$$
\stM\rho\stM = \stM [ \strho, \stM] + \stM^2 \strho
$$
and the similar
$$
\stM\rho\stM = -[ \strho, \stM]\stM + \strho\stM^2 
\q.
$$
Combining these gives
\beq
\lbl{eq210}
\stM \strho \stM = 
\frac{1}{2} \{\stM^2 , \strho \}
+ \frac{1}{2}  [ \stM, [\strho, \stM ]]
\q.
\eeq 
Using \re{eq210} and the contextual value equation 
\re{eq110}, $A = \sum_j \alpha_j \stE_j = \sum_j \alpha_j M^2_j$, 
we can rewrite the numerator of \re{eq150} as
\begin{eqnarray}
\lbl{eq220} 
\mbox{numerator of \re{eq150}} &=& 
 \sum_j \alpha_j \tr [ \stM_j \stP_f \stM_j \rho ] \nonumber\\
&=& 
 \sum_j \alpha_j \tr [ \stP_f \stM_j \rho \stM_j ] \\
&=& 
 \frac{1}{2} \tr [ \stP_f \{ \stA, \strho \}  ] 
+ \sum_j \frac{1}{2}\alpha_j \tr [ \stP_f [\stM_j ,  [\rho, \stM_j]\, ] 
\nonumber
\q.  
\end{eqnarray}
After division by the denominator of $\re{eq150}$, the first term 
gives DAJ's claimed (7) in the limit $g \goesto 0$, our \re{eq170}, 
and the second term gives
\begin{eqnarray}
\lbl{eq230}
\mbox{difference between weak limit of (6) and (7) =}
&& \nonumber \\
&& \nonumber \\
\lim_{g \goesto 0}
\frac{\sum_j \frac{1}{2}\alpha_j(g) \tr 
[ \stP_f [\stM_j (g) ,\,  [\rho, \stM_j (g)]\, ] 
\,]
}
{\tr [ \stP_f \strho ]}
.  
\end{eqnarray}
We shall call \re{eq230} the ``anomalous term''.
Since there is no obvious control over the size of the $\alpha_j (g)$, 
a counterexample is expected, but was surprisingly hard to find.

The Version 1 counterexample for the quoted claim of DAJ and the newer
counterexample for the new claim of DJ  are identical up to this point.
The difference is that the Version 1 counterexample used three $2 \times 2$
diagonal matrices as measurement operators, resulting in  a contextual value equation
\re{eq110} with multiple solutions, whereas the newer counterexample
uses three $3 \times 3$ diagonal matrices for which there is a unique
solution to \re{eq110}.   The newer counterexample could supercede the 
Version 1 example, but 
we retain the original counterexample because of its simple and 
intuitive nature (e.g., all steps can be mentally verified).

\subsection{The Version 1 counterexample }
The ``system observable'' $A$ for the counterexample will correspond
to a $2 \times 2$ matrix
\beq
\lbl{eq240}
A := \left[
\begin{array}{cc}
a & 0 \\
0 & b \\
\end{array}
\right]
\eeq
There will be three measurement operators: 
\begin{eqnarray}
\lbl{eq250}
 M_1 (g) &:=& 
\left[
\begin{array}{cc}
1/2 + g & 0 \\
0 & 1/2 - g
\end{array}
\right] 
, \ \ 
M_2 (g) := 
\left[
\begin{array}{cc}
1/2 - g  & 0 \\
0 & 1/2 + g 
\end{array}
\right] ,
\\
M_3(g) &:=& 
\left[
\begin{array}{cc}
\sqrt{1/2  - 2 g^2} & 0 \\
0 & \sqrt{1/2 - 2 g^2} 
\end{array}
\right] .
\nonumber
\end{eqnarray} 
Note that $M_3 (g)$ is uniquely defined by the
measurement operator equation $\sum_{j=1}^3 M^2_j (g) = 1$ and that
all three measurement operators approach multiples of the identity
as $g \goesto 0$, which assures weakness of the measurement.
Note also that $M_3 (g)$ is actually a multiple of the identity for 
all $g$, so the
commutators in the expression \re{eq230} for the anomalous term which 
involve $M_3$ vanish.  That is, $M_3$, and hence $\alpha_3$, 
make no contribution to the anomalous term. 

Writing out the contextual value equation \re{eq110} in components 
gives two scalar equations in three unknowns:
\begin{eqnarray}
\lbl{eq260}
(1/2 + g)^2 \alpha_1 (g) + (1/2 - g)^2 \alpha_2 (g) + 
(1/2 - 2g^2) \alpha_3 (g) &=& a \\
(1/2 - g)^2 \alpha_1 (g) + (1/2 + g)^2 \alpha_2 (g)  
+ (1/2 - 2g^2) \alpha_3 (g) &=& b \nonumber \q.
\end{eqnarray}
The solution can be messy because of the algebraic
coefficients.  However, for the case $a = 1 = b$, a solution can be 
obtained without calculation. 
This choice
of $a$ and $b$ corresponds to the system observable being the identity 
operator, so the measurement is not physically interesting, 
but it gives a mathematically valid example with minimal calculation.
Later we shall indicate how counterexamples can be obtained 
for other choices of $a$ and $b$ from appropriate solutions of \re{eq260}.

Assuming $a = 1 = b$, the system \re{eq260} can be rewritten 
\begin{eqnarray}
\lbl{eq270}
(1/2 + g)^2 \alpha_1 (g) + (1/2 - g)^2 \alpha_2 (g)  
&=& 1 - (1/2 - 2g^2) \alpha_3 (g)  \\ 
(1/2 - g)^2 \alpha_1 (g) + (1/2 + g)^2 \alpha_2 (g)  
 &=& 1 -  (1/2 - 2g^2) \alpha_3 (g) \nonumber 
\q.
\end{eqnarray}
We will think of $\alpha_3 (g)$ as a free parameter to be arbitrarily chosen,
and as noted previously, the choice will not affect 
the anomalous term \re{eq230}.  

Viewed in this way, \re{eq270}
becomes a system of two linear equations in two unknowns 
which become the same equation if 
$\alpha_2 =  \alpha_1$, with solution 
\beq
\lbl{eq280}
\alpha_2(g) =\alpha_1 (g) = 
\frac{1 - (1/2 -2g^2)\alpha_3(g)}
{ (1/2 +g)^2 + (1/2 - g)^2} = 
\frac{1 - (1/2 -2g^2)\alpha_3(g)}{1/2 + 2g^2)}
.
\eeq 

Since $\alpha_3$ can be chosen arbitrarily, also $\alpha_2 = \alpha_1$ can 
be arbitrary; we shall choose $\alpha_3(g)$ so that 
\beq
\lbl{eq290}
\alpha_2(g) = \alpha_1 (g) = \frac{1}{g^2}
\q.
\eeq 
To see that this solution 
will produce a counterexample,
note that for  
$$ 
\rho = 
\left[
\begin{array}{cc}
\rho_{11} & \rho_{12}\\
\rho_{21}   & \rho_{22}
\end{array}
\right]
$$
and for any diagonal matrix 
$$
D = 
\left[
\begin{array}{cc}
d_{1} & 0\\
0   & d_2
\end{array}
\right]
, 
\q [D, \rho] = 
\left[
\begin{array}{cc}
0  & (d_1 - d_2) \rho_{12}\\
(d_2 - d_1)\rho_{21}   & 0 
\end{array}
\right], 
\  \mbox{and}
$$
$$
[\, D, [D, \rho]\,] = 
\left[
\begin{array}{cc}
0  & (d_1 - d_2)^2\rho_{12}\\
(d_2 - d_1)^2 \rho_{21}   & 0
\end{array}
\right]
.
$$
In particular for $j = 1,2$,
$$
[\ M_j(g),\, [M_j(g), \rho]\,] =  
\left[
\begin{array}{cc} 
0 & 4g^2 \rho_{12} \\
4g^2 \rho_{21} &  0
\end{array}
\right] \q,
$$
and since $M_3 (g)$ is a multiple of the identity, $[M_3(g), \rho] = 0.$
Hence \re{eq230} becomes:
%
\beq
\lbl{eq295}
\frac{-(1/2)\tr [P_f \sum_j  \alpha_j 
 [M_j(g),\, [M_j(g), \rho]\ ]\ ]}
{\tr [ \stP_f \strho} 
=
\frac{
- \tr [P_f 
\left[
\begin{array}{cc}
0 & 4\rho_{12}\\
4\rho_{21}   & 0\\ 
\end{array}
\right] ]
}
{\tr [ \stP_f \strho] }
\q. 
\eeq
This is easily seen to be nonzero for $\rho_{12} \neq 0$ and appropriate
$P_f$.  For a norm 1 vector $f := 
(f_1, f_2) $  
\beq
\lbl{eq300}
\mbox{weak limit of (6)} = 
\frac{\tr [ P_f \{A, \rho \} ]}
{2 \tr [P_f \rho ]} 
+ \frac{- 8 \Re ({f}^*_2 f_1 \rho_{21}) }
{|f_1|^2 \rho_{11}  + 2 \Re ({f}^*_2 f_1 \rho_{21}) + |f_2|^2 \rho_{22} }.  
\eeq

The counterexample just given assumed that the system observable 
$A := \mbox{diag$\{ a,b\} $ } $
was the identity to make the calculations easy, but counterexamples
can be obtained for any system observable.  For example, if $A$ is the 
one-dimensional projector 
$A := \mbox{diag$\{1,0\} $ } $, and if system \re{eq270} is solved with
$\alpha_1(g) := 1/g^2$, then $\alpha_2(g) = 1/g^2 - 1/(2g)$, and  
the weak limit of the anomalous term is the same as just calculated for
$A = I$. \cite{parrott3} 

DJ \cite{DJ} adds additional (very strong) hypotheses to those of DAJ
which the counterexample just given does not satisfy.%
\footnote{  
However, the fact that these additional conditions cannot reasonably be
inferred from DAJ is not made clear by DJ, and a casual reader might well
obtain the opposite impression.
} 
Assuming these
additional conditions, DJ attempts to prove that (6) evaluates to (7)
in their ``minimal disturbance limit''. 
The next sections will present a more powerful counterexample which 
disproves this new claim of DJ.   

Originally a new paper with the more powerful counterexample was 
submitted to the arXiv, but a moderator rejected it.  He thought that
instead, Version 1 should be modified.
Rather than waste time on a 
lengthy and unpleasant appeal, I decided that it would be easier to do that.

The paper to this point is Version 1.  
The sections following comprise essentially
the rejected arXiv paper, which presents the more powerful counterexample
and critically analyzes DJ.  

The new counterexample is fairly simple,
utilizing three $3 \times 3$ matrices, 
but not as intuitive as one would like.  It was found by analyzing the 
properties that measurement operators might have in order that (7) could
be shown false, and then playing with parametrized $3\times 3$ 
measurement operators, trying to adjust the parameters so that (7) would 
not hold.  The Version 4 counterexample is simpler and more powerful than
the Version 2 counterexample.  
No doubt even simpler counterexamples could be found. 
Besides the new counterexample, we attempt to clarify some statements in DJ
which we think might be misleading. 
\section{The new additional hypotheses for the claim that (6) implies (7)
in the 
``minimal disturbance limit''}
Section 5 of DJ lists several additional assumptions, the most important
of which are:
\begin{enumerate}
\item
The $M_j$ commute with each other and $A$ (so they can all be represented
by diagonal matrices).    
\s
This is a strong assumption.
It is hard to imagine how it could reasonably be inferred 
or even guessed from DAJ.  The closest reference in DAJ to something similar 
is the following.
\begin{quote}
``To {\em illustrate} [emphasis mine] the construction of 
the least redundant set of [contextual values],
we consider the case when $\{\dajM_j \}$ and $\dajA$ all commute.''
\end{quote}
Nothing is said about this being a general assumption for the rest of the
paper.  Indeed, such an assumption would seriously restrict the applicability 
of the following definition (6) of ``general conditioned average''
$_f \lb \dajA \rb $, which requires no such assumption.
I studied DAJ for months without ever being led to even consider 
the possibility that this might be an 
{\em assumption} for the general claims of its abstract.
\item
The contextual values $\vec{\alpha} = (\alpha_1, \ldots, \alpha_N)$ are
obtained from the eigenvalues $\vec{a} = (a_1, \ldots, a_m)$  of
$A = \mbox{diag} ( a_1, \ldots , a_m)$ as
$$
\vec{\alpha} = F^{(+)} \vec{a}
$$
where $F$ is an $N \times m$ matrix satisfying $F \vec{\alpha} = 
\vec{a}$ and $F^{(+)}$ its Moore-Penrose pseudo-inverse.
The Version 1 counterexample does not satisfy this condition.
\s
Relying only on what is written in DAJ, 
it would be very hard for a reader to guess that this is supposed to be 
a {\em hypothesis} for (6), or for a claim that (6) implies (7), or both. 
(I did consider these possibilities, but rejected them as too unlikely,
as will be explained later in more detail.)
The only  
passage of DAJ which seems possibly relevant is:
\begin{quote}
``$\ldots$ we propose that the physically sensible choice of 
[contextual values] is the least redundant set uniquely related to the 
eigenvalues through the Moore-Penrose pseudoinverse.'' 
\end{quote}
DAJ gives no reason why this should be the ``physically
sensible choice''.   (DJ does attempt to address this issue, but 
unconvincingly and badly incorrectly, as will be discussed later.)  
Again, to assume this
would seem to artificially limit the applicability of (6), 
since (6) is correct independently of 
this assumption.   
\end{enumerate}
We do not list the other hypotheses for DJ's attempted proof that (6)
implies (7) because they are more technical and less surprising 
than the two just discussed.  
Our counterexample will satisfy all of the hypotheses listed in DJ.  
The counterexample for Version 2 has been replaced by a 
simpler example in Version 4.  
\section{A counterexample to the claim of DJ}

Section  V  of DJ entitled ``General Proof'' attempts to show that
(6) implies (7) under their listed hypotheses.  The present section 
presents a counterexample which satisfies all of their listed  
hypotheses, yet the weak limit of their ``general conditioned average'' (6) 
is {\em not} the ``quantum weak value''(7). 

We follow identically the analysis of Section 3 
through equation \re{eq230}.  
This time, we use a system observable
\setcounter{equation}{200}
\beq
\lbl{eq500}
A = 
\left[
\begin{array}{ccc}
1 & 0 & 0 \\
0 & 0 & 0 \\
0 & 0 & 0 
\end{array}
\right]
\q.
\eeq
and three measurement operators
which are $3 \times 3$ diagonal matrices: 
\begin{eqnarray}
\lbl{eq510}
M_1 (g) &:=& 
\left[
\begin{array}{ccc}
\sqrt{1/2 + g} & 0 & 0 \\
0 & \sqrt{1/2} & 0 \\
0 & 0 & \sqrt{1/2 + g} 
\end{array}
\right]\q, \nonumber\\
M_2(g) &:=& 
\left[
\begin{array}{ccc}
\sqrt{1/3 + g^2} & 0 & 0 \\
0 & \sqrt{1/3 + g} & 0 \\
0 & 0 &  \sqrt{1/3} 
\end{array}
\right]\q, \nonumber\\
\q M_3(g) &:=& 
\left[
\begin{array}{ccc}
\sqrt{1/6 - g - g^2} & 0 & 0 \\
0 & \sqrt{1/6 - g } & 0 \\
0 & 0 & \sqrt{1/6 - g} 
\end{array}
\right].
\end{eqnarray}
The  contextual values $\vec{\alpha} = (\alpha_1, \alpha_2,
\alpha_3)$ satisfy $ F \vec{\alpha} = \vec{a} := (1,0,0)^T$ with
\beq
\lbl{eq520}
F =  
\left[
\begin{array}{ccc}
1/2 + g &  1/3 + g^2 &  1/6 - g - g^2\\
1/2  &  1/3 + g & 1/6 - g \\
1/2 + g & 1/3 & 1/6 - g 
\end{array}
\right]
\eeq
The matrix $F$ is invertible  with inverse (which is also equal to the 
Moore-Penrose pseudoinverse $F^{(+)}$)
\beq
\lbl{eq525}
F^{(+)} = F^{-1} = 
\mbox{\Large
$
\left[
\begin{array}{ccc}
\frac{1-6g}{6g^2} & \frac{1-2g}{2g} & \frac{-1+9g}{6g^2} \\
\frac{1-6g}{6g^2} & \frac{1 + 2g}{2g} & \frac{-1+3g}{6g^2} \\
\frac{-5 - 6g}{6g^2} & \frac{1+2g}{2g} & \frac{3g+5}{6g^2} 
\end{array}
\right].
$
}
\eeq
The important thing to note is that the first column, which is also
$(\alpha_1, \alpha_2, \alpha_3 )^T$,  is of leading order $1/g^2$
as $ g \goesto 0$, which is all that the subsequent proof will use: 
\beq
\lbl{eq527}
\alpha_1 (g) = \alpha_2 (g) = \frac{1-6g}{6g^2} ,\q 
\alpha_3 (g) = \frac{-5 - 6g}{6g^2}.
\eeq
The full inverse \re{eq525} was obtained from a computer algebra program, 
and the first column (which is all that the counterexample will use) 
was also checked by hand using Cramer's rule. 

Equations \re{eq190} through \re{eq230} write 
the ``general conditioned average'' $_f\lb A \rb$ of (6) 
as a sum of two terms, one of which
evaluates to (7) in the weak limit $g \goesto 0$.  The other term, called
the ``anomalous term'', is given by \re{eq230} as:
\begin{eqnarray}
\lbl{eq530}
\mbox{difference between weak limit of (6) and (7) =}
&& \nonumber \\
&& \nonumber \\
\lim_{g \goesto 0}
\frac{\sum_k \frac{1}{2}\alpha_k(g) \tr 
[ \stP_f [\stM_k (g) ,\,  [\rho, \stM_k (g)]\, ] 
\,]
}
{\tr [ \stP_f \strho ]}
.  
\end{eqnarray}
To disprove the claim of DJ, we need to show that there exists
a state $\rho$ and vector $f$ such that the anomalous term 
does not vanish.   

It is well-known that the only  matrix $S$ such that for all projection
matrices $P_f$, $\tr [ P_f S] = 0,$ is the zero matrix $S=0$.%
\footnote{A computational proof is routine, but to see this without
calculation, recall that $\lb S, T \rb := \tr S^\dag T $ defines a 
complex Hilbert space structure 
(i.e., positive definite complex inner product) 
on the set of $n \times n$ matrices.  
If $\lb S, T \rb$ 
vanishes for all projectors $T = P_f$, then (by the spectral theorem),
it vanishes for all Hermitian $T$, and hence for all $T$, in which case  
$S$ is orthogonal to all elements of this Hilbert space and hence is the
zero element.
}
Hence it will be enough to show that 
\beq
\lbl{eq540}
\lim_{g \goesto 0}
\sum_k -\frac{1}{2}\alpha_k(g) 
 [\stM_k (g) ,\,  [ \stM_k (g), \rho ] \, ] 
\neq 0.
\eeq
for some mixed state $\rho$ such that for all nonzero vectors $f$,
$\tr [P_f \rho] \neq 0$.

First note that for any diagonal matrix $D = \mbox{diag} (d_1, d_2, d_3)$
and any matrix $\rho = (\rho)_{ij}$,
\beq
\lbl{eq550}
[D,[D,\rho]]_{ij} = (d_i - d_j)^2 \rho_{ij} \q.
\eeq
In the cases of interest to us, $D$ will be one of the measurement operators
  $M_k (g)$ , $ (d_i(g) - d_j (g))^2 = O(g^2)$ for all $i,j$, and for some
$i,j,$ the leading order of $(d_i (g) - d_j (g))^2$ is actually $g^2.$ 
The $\alpha_k (g)$ 
all diverge like
$1/g^2$ as $g \goesto 0$.  Thus we can see without calculation that 
we will obtain a counterexample unless some unrecognized relation forces
the terms of \re{eq540} to exactly cancel.%
\footnote{ 
One useful observation that we can make 
from what we have done so
far without detailed calculation 
is that the attempted proof of DJ is likely wrong or at least 
seriously incomplete, since that attempted proof concludes the vanishing 
of \re{eq540}
on the basis of order of magnitude arguments only.  Though framed in 
different language, it essentially says that \re{eq540} must vanish because
they think that
$\alpha_k (g) = (F^{(+)} (g) (1,0,0)^T)_k = O(1/g)$ 
(contradicting \re{eq527}), 
while $[M_j(g), [M_j(g), \rho]] = O(g^2).$
}

That cancellation does not occur in this case can be seen with
minimal calculation
as follows.  In \re{eq550}, take $(i,j) := (1,2)$, and note that 
$(d_1 - d_2)^2$ is always non-negative.  When $D = M_3 (g)$, from the 
power series 
$$
\sqrt{c + x} = \sqrt{c} + \frac{x}{2\sqrt{c}} + O(x^2)\q,
$$ 
one sees that $(d_1 - d_2)^2 = O(g^4)$, and since $\alpha_3(g)$ is  
only $O(g^{-2})$, the $k=3$ term in \re{eq540} vanishes in the limit 
$g \goesto 0$.  

We also have 
$$
\alpha_1 (g) = \alpha_2 (g) = \frac{1}{6g^2} - \frac{1}{g} 
 \q, 
$$
and for either $D = M_1(g)$ or $D = M_2(g)$,
$$
(d_1 - d_2)^2 = (g/\sqrt{2})^2 + O(g^2))^2 = g^2/2 + O(g^3) \q
$$
So, in the limit $g \goesto 0$, \re{eq540}
evaluates to 
\beq
\lbl{eq552}
-\frac{1}{2}  \frac{1}{6} \frac{1}{2} \rho_{12} = -\frac{\rho_{12}}{24} \q.
\eeq
Note that all we care about is that \re{eq552} does not always vanish,
and this can be seen solely from the fact that  
$\alpha_1$ and $\alpha_2$ have the same sign, so that the $k=1,2$
terms in \re{eq540} are negative multiples of $\rho_{12}$ 
which do not vanish identically in the limit $g \goesto 0$. 

To finish the proof, let $\rho$ be a positive definite state 
(i.e., all eigenvalues
strictly positive) such that $\rho_{12} \neq 0$.  Such a state can 
be constructed by starting with a positive definite diagonal state and adding
a small perturbation to assure $\rho_{12} \neq 0$ (which will result in
a positive definite state if the perturbation is small enough).
Since $\rho$ is positive definite, $\tr[\rho P_f] \neq 0$ for all nonzero
vectors $f$, and we are done.

\section{Discussion of DJ}
\subsection{Possible error in DJ's proof}
The counterexample given above unfortunately relies on some detailed calculation.
A conceptual counterexample would certainly be preferable.  A reader
interested in discovering the truth of the matter will be faced with the 
unpleasant choice of wading through DJ's dense proof or checking the boring
details of the counterexample.  For such readers, it may be helpful if
we point out what seems a potentially erroneous step in DJ's proof.

A step which caused me to question their proof occurs at the very end of
their Section V: 
\begin{quote}
``$\ldots$ to have a pole of order higher than $g^n$ [$n=1$ in the 
counterexample] then there must be at least one relevant singular value 
with an order greater than $g^n$.
[The counterexample has a singular value of order $g^2$.]
However, if that were the case 
then the expansion of $F$ to order $g^n$ would have a relevant singular
value of zero and therefore could not satisfy (25) $\ldots$''
\end{quote}
I have not been able to guess a meaning for 
``the expansion of $F$ to order $g^n$ would have a relevant singular
value of zero'' under which the last sentence would be true.
\subsection{Significance of the Moore-Penrose pseudo-inverse}
The original paper DAJ \cite{DAJ} introduced the Moore-Penrose pseudo-inverse
as follows:
\begin{quote}
``$\ldots$ we propose that the physically sensible choice of [contextual
values $\vec{\alpha}$] is the least redundant set 
uniquely related to the eigenvalues [$\vec{a} = (a_1, \ldots, a_m)$ with 
$ A = \mbox{diag} (a_1, \ldots, a_m$)]
through the Moore-Penrose pseudoinverse.''
\end{quote}
I puzzled for a long time over this statement.  Besides the fact that 
the meaning of ``least redundant set'' was obscure to me, they give no
reason why this choice (which presumably means 
$\vec{\alpha}  = F^{(+)} \vec{a}$,
with $F^{(+)}$ the Moore-Penrose pseudo-inverse) should be considered 
the unique ``physically sensible'' choice, or even {\em a} physically
sensible choice.  The arXiv paper DJ which we are discussing attempts
to fill this gap, but the attempt relies on erroneous mathematics and 
is unconvincing. 

Before starting the discussion of this attempt, let me remark that
although the attempt seems partly aimed at 
invalidating the counterexample of \cite{parrott4}, 
it is basically irrelevant to that aim.  That 
counterexample is a valid mathematical counterexample to 
a mathematical claim of DAJ as I imagine the vague exposition of 
DAJ would probably be interpreted by most readers.  
Though the counterexample
uses a particular solution of the contextual value equation 
$F \vec{\alpha} = \vec{a}$, it was never claimed that this solution 
has any physically desirable properties.  
Though DJ does {\em not} show that the counterexample is unphysical
as DJ claims, even if it were shown unphysical, it would still disprove
the claim that (6) necessarily evaluates to (7) 
in the ``minimal disturbance limit''.  A reader of DAJ cannot reasonably
be expected to guess that the definition of ``minimal disturbance limit''
is supposed to include the pseudo-inverse prescription. 

Therefore, the discussion will be directed toward analyzing the claim
of DJ that the pseudo-inverse solution should be preferred because 
DJ thinks (incorrectly) that 
\begin{quote}
``$\ldots$ the pseudo-inverse solution will choose the solution that
generally provides the most rapid statistical convergence for observable
measurements on the system.'' 
\end{quote}
A careful analysis of DJ's  
argument for this claim will reveal flaws which invalidate it.

DJ writes:
\begin{quote}
``With the pseudo-inverse in hand, we then find a uniquely specified
solution $\vec{\alpha}_0 = F^{(+)} \vec{a}$ that is directly related
to the eigenvalues of the operator.  Other solutions 
$\alpha = \vec{\alpha}_0 + x $ of (3) will contain additional components
in the null space of $F$, and will thus deviate from this least 
redundant solution. [True if sympathetically interpreted, but tautological.] 
Consequently, the solution $\alpha_0$ has the least norm of all solutions
$\ldots$''
\end{quote}
The Euclidean norm $||\vec{\alpha}||^2 := \sum_j \alpha^2_j$ in the real Hilbert space
$R^n$ has no physical significance in quantum theory.  Why is it relevant
that $\vec{\alpha}_0$ has least norm?  The discussion immediately following may
possibly be intended to answer this, but when analyzed it only tautologically
repeats what has already been said.  However, an inattentive reader could
easily get the impression that something had been proved. 

DJ thinks that this immediately following discussion (at the bottom of
the first column of p.4) gives ``mathematical reasons for using the
pseudoinverse'', but in fact no convincing reason has been given.

The next paragraph continues:
\begin{quote}
``In addition to the mathematical reasons for using the pseudoinverse in this
context [referring to the discussion just analyzed, which doesn't give any
convincing mathematical reason], there is an important physical one that
we will now describe.  As mentioned, a fully compatible detector 
can be used together with the contextual values to reconstruct
any moment of a compatible observable.
However, since the detector outcomes are imperfectly correlated with
the observable, the contextual values typically lie outside of the 
eigenvalue range and many repetitions of the measurement must be practically
performed to obtain adequate precision for the moments.  
{\em Importantly,
the uncertainty in the moments is controlled by the the variance, 
not of the observable operator, but of the contextual values themselves.}
[emphasis mine]''
\end{quote}
Consider a probability space with outcomes 
$\{1, 2, \ldots , n\}$ with probability  $p_j $ for outcome $j$.  
A  {\em random variable} $v$ is an assignment $j \mapsto v_j$ of a 
real number $v_j$ to each outcome $j$.  The {\em mean} $\bar{v}$
of $v$ is defined as usual by 
$$
\bar{v} := \sum_j v_j p_j \q,
$$ 
and the {\em variance} $\tau^2$ of $v$ is defined by 
$$
\tau^2 := \sum_j (v_j - \bar{v})^2 p_j = \sum_j v_j^2 p_j - \bar{v}^2 \q.
$$
Here we use the symbol $\tau^2$ instead of the customary $\sigma^2$ to denote
the variance to avoid confusion with the different $\sigma^2$ defined in DJ 
(as the second moment).

One can speak of the 
``variance'' of a random variable on a classical probability space, or 
of the ``variance'' of quantum observable measured in a given state.  
But what can it mean to speak of the ``variance'' of contextual values
$\alpha_j$?  Contextual values are 
are {\em predefined} to 
satisfy
$$
A = \sum_j \alpha_j M_j^\dag M_j
\q, \eqno (\ref{eq110})
$$ 
where $A$ is the ``system observable'' and $\{M_j\}$ a collection of
measurement operators.  What is {\em measured} are the outcomes $j$.  

However, even though we know the contextual values beforehand from 
\re{eq110}, one might speak of ``measuring'' them 
in the following sense. To every outcome $j$ corresponds a contextual
value $\alpha_j$.  A given state of the system $\rho$ 
makes the set of all outcomes $j$ into a probability space by 
assigning a probability $p_j$ to each outcome $j$:  
$p_j = \tr [M^\dag_j M_j\rho] $.
The assignment $j \mapsto \alpha_j$
is a random variable on this probability space, and it is meaningful
to speak of its ``variance''.  
The subsequent analysis assumes 
that this is the meaning that DJ intended.  This discussion may seem
inappropriately elementary, but I was initially puzzled about this point, and
it cannot hurt to make it explicit.  

Note that the mean $\bar{\alpha} = \tr [A\rho]$ of this random variable 
is the same no
matter how the contextual values $\alpha_j$ are chosen so long as they 
satisfy the contextual value equation \re{eq110}.
That implies that choosing the contextual values so as to minimize the
true variance $\tau^2$ in a given state is equivalent to minimizing
the second moment $\sigma^2$. 
Note also that the mean and variance implicitly depend on the state $\rho$,
and that there is no reason to think that one might be able to choose the
contextual values so as to minimize the variance in {\em all} states.  

DJ continues:
\begin{quote}
``Consequently, it is in the experimenters best interests to minimize the
second moment of the contextual values,
\beq
\lbl{eq610}
\sigma^2 = \sum_j \alpha^2_j p_j,
\eeq
where $p_j$ is the probability of outcome $j$.''
\end{quote} 
DJ correctly identifies $\sigma^2$ as the second moment, but 
unless read very carefully, 
the subsequent discussion 
could encourage confusion of $\sigma^2$ with the true 
variance $\tau^2$.   

Next DJ notes that $|| \vec{\alpha} ||^2$ is a (very crude) upper bound 
for $\sigma^2$: 
$$ 
\sigma^2 := \sum_j \alpha^2_j p_j \leq \sum_j \alpha^2_j = 
||\vec{\alpha}||^2 \q. \eqno (*)
$$ 
\begin{quote}
``In the absence of prior knowledge about the system one is dealing with,
this is the most general bound one can make.  
Therefore, the pseudo-inverse solution will choose the solution that
generally provides the most rapid statistical convergence for observable
measurements on the system.'' 
\end{quote} 
This is highly questionable.
Although it may not be clear at this point, subsequent
paragraphs make clear that DJ is claiming that 
it is legitimate to use $||\vec{\alpha}||^2$ as a sort of estimate 
for $\sigma^2$, the strange and invalid justification for the claim 
being the sentences
of the quote following equation (*).%

DJ's next paragraph computes $|| \vec{\alpha} (g) ||^2$ for both the $\vec{\alpha}(g)$
used in the counterexample of \cite{parrott4} and for the 
pseudo-inverse solution $\vec{\alpha}_0 (g) = F^{(+)}(g) \vec{a}$,
using $|| \vec{\alpha} (g) ||^2$ as a kind of crude estimate
for $\sigma^2 = \sigma^2 (g) = \sigma^2 (g, \rho)$. 
\begin{quote}
``For the case of the counterexample, the Parrott solution (13) 
[(13) should be (11)] has to leading order the bound on the  variance 
$$
||\vec{\alpha}||^2 = \frac{3}{g^4} - \frac{3(a-b)}{2g^3} + 
O(\frac{1}{g^2}), \eqno (15)
$$
while the pseudoinverse solution (11) [(11) should be (13)] has to leading
order the bound
$$
||\vec{\alpha}||^2 = \frac{(a-b)^2}{8g^2} + \frac{2}{3}(a+b)^2
+ O(g^2).  \eqno (16) 
$$ 
For any observable $\vec{a}$, {\em the Parrott solution has detector 
variance  of order $O(1/g^4)$}[emphasis mine], which would swamp
any attempt to measure an observable near the weak limit.  $\ldots$
However, the pseudoinverse solution has a detector variance of order
$O(1/g^2)$ in the worst case; $\ldots$ '' 
\end{quote} 
What invalidates the argument is the use 
of the crude upper bound (*) as an estimate for the second moment 
$\sigma^2$ and the subsequent claim that ``the Parrott
solution has detector variance of order $O(1/g^4)$
$\ldots$ ''.%
\footnote{DJ incorrectly identifies $\sigma^2$ as the ``variance'' instead 
of the second moment, but this is a mere slip.   
Ignoring this slip, technically 
one could argue that this statement is correct
because to say that a quantity is $O(1/g^4)$ only means that it increases
{\em no faster} than $1/g^4$ as $g \goesto 0$.  
For example, $g^8 = O(1/g^4)$.  However,
in the context and taking into account the typically sloppy use 
of the ``big-oh'' notation in the physics literature, most readers would
probably interpret this passage as claiming that the ``Parrott solution''
has variance of {\em leading order} $1/g^4$, which would be an invalid 
conclusion from the argument.
}

Solely from {\em upper bounds} for two quantities, one cannot
draw any reliable conclusions about the relative size of the 
quantities themselves.  
To see this clearly in a simpler context which uses essentially 
the same reasoning,
consider the upper bounds
$$ x < x^4 \q \mbox{and} \q  x^2 < x^3 $$
for real numbers $x > 1$.  
From the fact that the first upper bound $x^4$ (for $x$) is larger than
the second upper bound $x^3$ (for $x^2$), we cannot conclude that 
$x$ is larger than $x^2$ for $x > 1$. Yet DJ's argument relies on
this type of incorrect reasoning.  

In the interests of following closely the exposition of DJ, we passed 
rapidly over (*).  Let us return to analyze it more closely: 
$$ 
\sigma^2 := \sum_j \alpha^2_j p_j \leq \sum_j \alpha^2_j = 
||\vec{\alpha}||^2 \q. \eqno (*)
$$ 
\begin{quote}
``In the absence of prior knowledge about the system one is dealing with,
this is the most general bound one can make.  
Therefore, the pseudo-inverse solution will choose the solution that
generally provides the most rapid statistical convergence for observable
measurements on the system.'' 
\end{quote} 
Note once again that $\sigma^2 = \sigma^2 (\rho)$ depends implicitly on
the state $\rho$ because the probabilities $p_j = \tr [\rho M^\dag_j M_j]$ 
of outcome $j$ depend on $\rho$.  Keeping this in mind, one sees how
crude the upper bound (*) really is.  

For a nonzero system observable $A$, equality holds in (*) 
(i.e., $\sigma^2 (\rho) = ||\vec{\alpha}||^2)$ only in the trivial case 
in which one particular $p_J = 1$ and the others vanish, and in addition,
$\alpha_j (g)  =0$ for $j \neq J$.  That corresponds to the trivial case in 
which there is effectively only one measurement operator $M_J (g)$ 
satisfying $\alpha_J(g) M^\dag_J (g) M_J(g) = A$. 
(The other measurement operators play the role
of assuring that $\sum_j M^\dag_j M_j = I$, but do not contribute to the
estimation of the expectation of $A$ in the state $\rho$, $\tr [A\rho]$.) 
Since one of DJ's hypotheses (which we did not discuss above) is that
$\lim_{g \goesto 0} M_j (g) $ is a  multiple of the identity for
all $j$, also the system observable $A$  is a multiple of the identity.

The statement following (*), that ``this is the most general bound one can 
make'', seems a very strange form of reasoning.  Doubtless, (*) {\em was} the 
most general bound that the authors knew how to make, 
but it seems unscientific
to base an important argument on an unsupported personal belief 
that no one else can do better.  

In fact, 
a better bound is possible.  By the Cauchy-Schwartz inequality,
$$
\sigma^2 =  \sum_j \alpha^2_j p_j \leq [\sum_j (\alpha^2_j)^2]^{1/2} 
[ \sum_j p^2_j]^{1/2}
\leq   
\left [\sum_j \alpha^4_j \right]^{1/2}, 
\eqno (**)
$$
since $ 0 \leq p_j \leq 1$, so $p^2_j \leq p_j$ and 
$\sum_j p^2_j \leq \sum_j p_j = 1$.  That (**) is a better bound than (*)
when at least two $\alpha_j$ are nonzero
follows from 
$$   
\left[ \left[ \sum_j \alpha^4_j \right ]^{1/2} \right]^2 =  
\sum_j (\alpha^2_j)^2  <  \left[ \sum_j \alpha^2_j \right]^2 
= \left[ ||\vec{\alpha}||^2 \right]^2,
$$
because for any collection of at least two positive numbers 
$\{q_j\}$, $\sum q^2_j < [\sum_j q_j]^2 $.   

If the authors were to reformulate their proposal for the appropriate choice
of the contextual values in terms of this better bound,  
it seems unlikely that DAJ's proposed Moore-Penrose pseudo-inverse 
solution would minimize (**), or possible bounds even better than (**).
And as pointed out earlier, the physical meaning or appropriateness of 
minimizing a particular {\em upper bound} for the detector's second
moment remains obscure.  
\subsection{What is the ``physically sensible'' choice of contextual values?}

In many experiments (indeed, in all experiments known to me),
the system always starts in a known state $\rho$.  For such an experiment,
it seems to me that the ``physically sensible choice'' of contextual values
would be the choice that minimizes the detector variance 
$\tau^2 = \tau^2 (\rho)$ {\em in that initial state $\rho$}. 

It is a simple exercise to work out a necessary condition for this 
minimization, and the pseudo-inverse prescription does not necessarily 
satisfy it.  For the reader's convenience, we sketch the details.

The contextual values equation \re{eq110} for contextual values 
$\vec{\alpha} = (\alpha_1, \ldots, \alpha_N)$ can always be written as 
a linear system given by a vector equation
\beq
\lbl{eq800}
\vec{\alpha} = F(\vec{a}) \q,
\eeq
where $\vec{a}$ is a vector associated with the system observable $A$ and
$F$ a matrix whose size will depend on the dimension of $\vec{a}$.%
\footnote{If
$A$ or some of the measurement operators are not diagonal, 
then $\vec{a}$ will not be the vector of eigenvalues
of $A$ as in DJ.  For example, if $A$ is a general $2 \times 2$  
Hermitian matrix, then $\vec{a} = (a_1, a_2, a_3)$ 
may be taken to be the three-dimensional vector $(A_{11}, A_{12}, A_{22})$,
and in general, $\vec{a}$ may be formed from the components of $A$ on or 
above its main diagonal.}  
Given an initial state $\rho$, measurement operators $M_j$, and associated
probabilities $p_j = \tr [ \rho M^\dag_j M_j]$, we want to minimize 
the detector variance 
\beq
\lbl{eq810}
\tau^2 (\rho) := \sum_i p_i \alpha^2_i - \left( \sum_i p_i \alpha_i \right)^2 
\q.
\eeq
As noted in the preceding subsection, for a particular state $\rho$ and
taking into account the contextual value equation \re{eq110}, 
this is the same as minimizing 
the second moment
\beq
\lbl{eq820}
\sigma^2 (\rho) := \sum_i p_i \alpha^2_i \q. 
\eeq 
(To avoid confusion, we continue 
using DJ's nonstandard notation $\sigma^2$ for the second 
moment instead of the variance.)

Let $\vec{\alpha}^P$ denote a particular solution of 
$F(\vec{\alpha}) = \vec{a}$.  
Then the general solution of 
$F(\vec{\alpha}) = \vec{a}$ is $\vec{\alpha} = \vec{\alpha}^P + \vec{\eta}$ 
with
$\vec{\eta}$ in the nullspace Null($F$) of $F$, and    
\beq
\lbl{eq830}
\sigma^2 := \sum_i p_i \alpha^2_i = \sum_i p_i (\alpha^P_i)^2
+ 2 \sum_i p_i \alpha^P_i \eta_i + \sum_i p_i \eta^2_i .
\eeq
For small $\vec{\eta}$, a nonvanishing linear second term will dominate the 
quadratic third term,%
\footnote{More precisely, if for some $\eta$ the linear term does not 
vanish, then replacing $\eta$ by $x \eta$, with $x$ real,  gives a quadratic
function in $x$ with nonvanishing linear term, which cannot have a minimum
at $x = 0$.}   
and we see that if $\vec{\alpha}^P$ 
is to minimize $\sigma^2$,
then the vector 
$(p_1 \alpha^P_1, \ldots , p_N \alpha^P_N)$ 
must be orthogonal
to the nullspace of $F$.  This is the necessary condition mentioned earlier.

Thus it seems to me that a ``physically sensible'' choice of contextual 
values in this situation should satisfy this necessary condition.   
However, the pseudo-inverse solution is abstractly defined by the different
condition 
that $\vec{\alpha}^P = (\alpha^P_1 , \ldots , \alpha^P_N)$ 
be orthogonal to Null($F$).%
\footnote{This is discussed but not proved in the Appendix to 
\cite{parrott3}.  A formal statement and proof can be found in 
\cite{campbell}, p.\ 9, Theorem 1.1.1. 
}  

Even if the state $\rho$ is not known from the start, to estimate 
the expectation of $A$ as $\sum_j \alpha_j \tr [ M^\dag_j M_j \rho ] = 
\sum_j \alpha_j p_j$, one needs to estimate
the $p_j$ as frequencies of occurence of outcome $j$, so the $p_j$ can be
regarded as experimentally determined to any desired accuracy.  Given these
$p_j$, one can then choose the solution $\vec{\alpha}$ to the contextual
value equation $F(\vec{\alpha}) = \vec{a}$ to minimize \re{eq830} and
the detector variance.  
This procedure for minimizing the 
detector variance will rarely result in the pseudo-inverse solution.
\subsection{Does DAJ assume that contextual values $\vec{\alpha}$
come from the Moore-Penrose pseudo-inverse, 
$\vec{\alpha} = F^{(+)}\vec{a}$?} 
We have seen that none of the reasons that DJ gives for determining contextual
values by the pseudo-inverse construction, 
\beq
\lbl{eq615}
\vec{\alpha} = F^{(+)}\vec{a}\q, 
\eeq
hold up under scrutiny. 
DAJ doesn't give any valid reasons, either.
Its ``general conditioned average'' (6) does not require this hypothesis, nor
the hypothesis that the system observable $A$ and measurement operators 
$M_j$ mutually commute.  Why assume something that is not needed?

DJ gives the false impression that DAJ unequivocally assumes \re{eq615}
as a hypothesis.  For example,   
\begin{quote}
``The problem with Parrott's counterexample is that he ignores this 
discussion [of defining the contextual values by the pseudo-inverse
prescription $\vec{\alpha} := F^{(+)} \vec{a}$] $\ldots$''. 
\end{quote}
The totality of this ``discussion''  is the single sentence:
\begin{quote}
``$\ldots$ we propose that the physically sensible choice of 
CV is the least redundant set uniquely related to the eigenvalues through
the Moore-Penrose pseudoinverse.''
\end{quote}

DAJ does devote
a long paragraph to a complicated method of defining and 
calculating the Moore-Penrose pseudo-inverse, but that has nothing to do with 
the reasons for using the pseudo-inverse in the first place.
A reference to a mathematical text would have sufficed and saved sufficient
space to have clearly stated their hypotheses for (6) and for the 
claimed implication that (6) implies (7) in their ``minimal disturbance 
limit''.  If the authors don't tell us,  
how can we poor readers possibly guess that the pseudo-inverse prescription 
\re{eq615} is assumed as a hypothesis
for (6) (if in fact it is, which to this day I don't know), or if not,
as a hypothesis for a section which follows (6), such as the ``Weak values''
section?   

When I wrote \cite{parrott4} giving the counterexample, I did consider
the possibility that DAJ might possibly be assuming the
pseudo-inverse solution, but rejected it as implausible.  This was 
partly because they had previously sent me an attempted proof that their (6)
implies (7) in their ``minimal disturbance limit'' which if correct 
(it wasn't) would have applied to {\em any} solution $\alpha$, not just
the pseudo-inverse solution.  (It also would have applied even if the 
measurement operators and system observable did not mutually commute.)
So, I knew to a certainty that when DAJ was submitted, there was no 
reason for the authors to have assumed the pseudo-inverse prescription.  

Also, in the sweeping claim of DAJ's abstract that their ``general conditioned
average'' (6) 
\begin{quote}
``$\ldots$ converges uniquely to the quantum weak value in the minimal
disturbance limit'',
\end{quote}
by no stretch of the imagination could the reader guess that the 
technical pseudo-inverse
prescription would be part of the definition of ``minimal disturbance limit''.
And if the prescription is not part of the definition of ``minimal disturbance
limit'', then to justify the claim, the prescription would have to be taken
as part of the definition of their ``general conditioned average'' (6).     
But the latter alternative would artificially limit the applicability of (6), 
since (6) is correct no matter how the contextual values are chosen (subject
to the contextual value equation \re{eq110}).
\subsection{Section VI of DJ:}
The last four paragraphs of Section VI of DJ (entitled ``Discussion'')  are 
misleading and in some ways incorrect.  The reasons are given
in Section 11.1 of \cite{parrott3} and will not be repeated here. 
\s
\section{Acknowledgments}
I was surprised to see in DJ the acknowledgment:  ``We acknowledge
correspondence with S.\ Parrott''.  That made me wonder if protocol required 
that I provide a similar acknowledgment.  And if so, what should it say?
Would it be proper to acknowledge negative contributions as well as 
positive ones, and if so should I?  If I didn't, how would I explain
why I didn't simply ask the authors about some of the questionable points
in DAJ?

The (nearly unique) positive contribution of the authors of DAJ to 
\cite{parrott3}, \cite{parrott4}, and the present work 
was to furnish their original argument
that (6) implies (7) in their ``minimal disturbance limit''. 
That argument 
 brought to my attention the decomposition
of equation \re{eq210}, which was part of their attempted proofs. 

That argument was definitely incorrect because I found a counterexample
to one of its steps.  I sent the counterexample to the authors in mid-February,
but they never acknowledged it.  I made several subsequent inquiries about
other points in DAJ, but all were ignored.  I have not heard from them
since February 19. (It is now June 23).  
(What little correspondence we did exchange was uniformly
courteous.)  That is why I was unable to clarify other vague points 
in DAJ such as for which results (if any) 
the pseudo-inverse solution was assumed
as a hypothesis.

I intend to
eventually post on my website, www.math.umb.edu/$\sim$sp , a complete account
of the strange aspects of this affair, which has been unique in my
professional experience.  It will raise questions about the editorial
practices of influential journals of the American Physical Society, among
other issues. 

DJ acknowledges that their work was supported by two grants, at least one 
of which was taxpayer-supported via the National Science Fountation.  The
present work was not supported by any grants, unless donation of the author's
time might be considered a kind of ``grant''.  

If so, it is a ``grant'' to society in general. 
I have spent months trying to unravel DAJ, mostly without any help.  
I submit this to the arXiv to save others similar time. 
It is painful to realize that I have largely wasted my time for a contribution
so small, but it is satisfying to hope that the time saved by others
may result in larger contributions than I could have made. 
\\[2ex]
{\bf Added in version 8:} Version 2 of \cite{DJ}, arXiv:1106.1871v2
replies to the present work.  It was 
was published in J.\ Phys.\ A: Math.\ Theor.\ {\bf 45} 015304.  The
published version will be called DJpub below. 
\s
I thank the authors for noting a typo in the definition of the $(3,3)$ 
entry of the $3 \times 3$
matrix $M_2 (g)$ in the Section 5 counterexample on p.11. The original
entry $1/3$ should have been $\sqrt{1/3}$, and this correction has 
been made in this Version 7.  The original
analysis assumed the correct value, so apart from this substitution,
no changes were necessary.
\s 
DJpub reinterprets (unjustifiably, in my view) one of the hypotheses of
\cite{DJ} and notes that the counterexample given above does not satisfy
the reinterpreted hypothesis.  
An analysis of DJpub has been posted in arXiv:1202.5604, and 
an abbreviated version has been under consideration by J.\ Phys.\ A for
over 10 months (as of this writing, October 14, 2012).     

\end{document}